\documentclass[useAMS,usenatbib,onecolumn]{mn2e}
\title[Scattering by a Gyrating Electron]
{Scattering of Low-Frequency Radiation by a Gyrating Electron}

\author[S. A. Petrova]{S. A. Petrova
\thanks{E-mail: petrova@ira.kharkov.ua}\\
Institute of Radio Astronomy, NAS of Ukraine, 4, Chervonopraporna
Str., 61002 Kharkov, Ukraine}

\begin{document}

\date{Received\dots}

\pagerange{\pageref{firstpage}--\pageref{lastpage}} \pubyear{2007}

\maketitle

\label{firstpage}

\begin{abstract}
The scattering of electromagnetic radiation by the particle
gyrating in an external magnetic field is considered. Particular
attention is paid to the low-frequency case, when the frequencies
of incident radiation are much less than the electron
gyrofrequency. The spectral and polarization features of the
scattering cross-section are analyzed in detail. It is found that
the scattering transfers the low-frequency photons to high
harmonics of the gyrofrequency, into the range of the synchrotron
emission of the electron. The total scattering cross-section
appears much larger than that for the particle at rest. The
problem studied is directly applicable to the radio wave
scattering in the magnetosphere of a pulsar. The particles acquire
relativistic rotational energies as a result of resonant
absorption of the high-frequency radio waves and concurrently
scatter the low-frequency radio waves, which are still below the
resonance. It is shown that the scattering can affect the radio
intensity and polarization at the lowest frequencies and can
compete with the resonant absorption in contributing to the
low-frequency turnover in the pulsar spectrum. Moreover, the
scattering can be an efficient mechanism of the pulsar high-energy
emission, in addition to the synchrotron re-emission of the
particles. Other astrophysical applications of the scattering by
gyrating particles are pointed out as well.
\end{abstract}


\section{Introduction}

The presence of an external magnetic field may substantially
affect the process of photon scattering off an electron. The
classical non-relativistic consideration of the magnetized
Thompson scattering has been done in
\citet*{c70,c71,gl71,bs76,v79,bm79a,bm79b}. It has been found that
the role of magnetic field is significant unless the photon
frequency substantially exceeds the electron gyrofrequency,
$\omega\gg\omega_G\equiv eB/mc$, and the magnetized scattering
cross-section is characterized by peculiar angular and frequency
dependencies as well as specific polarization signatures. The
fully relativistic treatment of the magnetic cross-section in
terms of quantum electrodynamics has been developed in
\citet{h79,mp83,dh86}, and useful approximations have been given
in \citet{x85,dh89,g00}. The relativistic effects become important
in extremely strong magnetic fields approaching the critical value
$B_{\rm cr}=4.413\cdot 10^{13}$ G defined as $\hbar\omega_G(B_{\rm
cr})\equiv mc^2$, and at frequencies roughly comparable with
$\omega_G(B_{\rm cr})$. Then the magnetized scattering exhibits
principally new features, such as resonances at high harmonics of
the gyrofrequency and the possibility of electron excitation to
higher Landau levels.

The relativistic regime may be applicable to close neighborhoods
of the neutron stars, whose surface magnetic fields are typically
$\sim 10^{12}$ G and in case of magnetars may be as large as $\sim
10^{15}$ G. The neutron stars have surface temperatures $T_{\rm s
}\sim 10^5-10^6$ K, and the thermal X-ray photons are scattered
off the primary particles, which are accelerated by the
rotation-induced electric field of the neutron star to the
Lorentz-factors $\gamma_{\rm p}\sim 10^6-10^7$. The resonant
Compton scattering of the thermal radiation of the neutron star
can be efficient \citep{bs76,x85,dh89} and is believed to have
important implications. The upscattered photons are capable of
producing the electron-positron pairs and may compete with the
curvature emission of the primary particles in controlling the
pair production cascade in the polar gap of a pulsar
\citep{sd94,s95,l96}. Thus, the resonant Compton scattering may
substantially affect the characteristics of the secondary pulsar
plasma \citep{ha01a,ha01b,ae02,hm02}. This process can also
account for the high-energy spectra of magnetars
\citep*{tlk02,lg06,rzlt07,ft07,bt07,bh07}. For a review of other
applications of the resonant Compton scattering to pulsars see,
e.g., \citet{hl06}.

The scattering of the non-thermal radio emission of pulsars off
the secondary plasma particles is an essentially distinct process
and is also of interest. The secondary electron-positron plasma
streams ultrarelativistically, with $\gamma\sim 10^2$, along the
open magnetic lines of a pulsar and ultimately leaves the
magnetosphere as a pulsar wind. The radio emission is believed to
originate inside the plasma flow deep in the magnetosphere. Hence,
on its way in the magnetosphere and beyond, pulsar radiation
passes through the plasma and is subject to scattering off the
plasma particles. Since for radio frequencies $h\nu/mc^2\ll 1$,
the quantum effects on the scattering are negligible and the
classical treatment is appropriate. In the vicinity of the
emission region, the photon frequency in the particle rest frame
is much less than the electron gyrofrequency, $\omega\gamma
(1-\beta\cos\theta)\ll\omega_G$ (here $\beta$ is the particle
velocity in units of $c$ and $\theta$ is the angle between the
photon wavevector and the particle velocity), i.e. the magnetic
field is strong enough to affect the scattering process. As the
magnetic field strength rapidly decreases with distance from the
neutron star, $B\propto R^{-3}$, in the outer magnetosphere the
waves suffer cyclotron resonance, and in the pulsar wind the
scattering is non-magnetic.

The scattering of pulsar radio emission in the magnetic and
non-magnetic regimes has first been considered in
\citet{bs76,wr78}. Because of extremely high brightness
temperatures of pulsar radiation, $T_B\sim 10^{25}-10^{30}$ K, the
induced scattering strongly dominates the spontaneous one and can
be efficient in both regimes. Further studies
\citep{lp96,p04a,p04b} have demonstrated that the magnetized
induced scattering can lead to substantial redistribution of
intensity in frequency and pulse longitude and thus can account
for various phenomena characteristic of the observed radio pulses.

Close to the neutron star surface, the magnetic field is strong
enough for any transverse momentum of the electrons to be almost
immediately lost via synchrotron re-emission. However, it is not
the case in the outer magnetosphere, where the radio waves meet
the condition of cyclotron resonance. Correspondingly, in the
resonance region the waves are subject to cyclotron absorption
rather than resonant scattering. At the conditions relevant to
pulsar magnetosphere, the process of cyclotron absorption not only
affects the radio wave intensity
\citep*{bs76,lp98,p02,melr_a,melr_b}, but also leads to a
substantial increase of the transverse momenta of the absorbing
particles \citep{bs76,lp98,p02,p03}. Since the pulsar radiation is
broadband, $\nu\sim n\cdot 10^7-n\cdot 10^{10}$ Hz, the resonance
region is sufficiently extended. The particles entering the
resonance region acquire relativistic gyration energies straight
near its lower boundary, in the course of absorption of the waves
with $\nu\geq 10^{10}$ G \citep{p02}. Then the lower-frequency
waves, $\nu\ll 10^{10}$ Hz, which are still below the resonance,
$\nu\gamma (1-\beta\cos\theta)\ll\nu_G$, are subject to the
magnetized scattering off the relativistically gyrating particles.
This process will be examined in detail in the present paper.

The scattering by a gyrating electron differs substantially from
that by an electron at rest. For the electron at rest, the
scattering to high harmonics holds only within the framework of
the relativistic treatment, in the magnetic fields close to the
critical value and at high enough frequencies
\citep{h79,mp83,dh86}. (Note that the relativistic formalism of
the scattering has been developed only for the case when the
electron is initially at the ground Landau orbital.) For the
gyrating electron, the high-harmonic scattering is a purely
classical effect and it may be efficient in arbitrary magnetic
fields for the incident frequencies below the electron
gyrofrequency. In application to pulsars, the scattering by the
electrons at high Landau levels transfers the radio photons into
the optical and X-ray ranges and thus provides a physical
connection between the radio and high-energy emissions of a
pulsar. Other astrophysical applications of the process, e.g., to
synchrotron sources, are not excluded as well.

It should be noted that the pulsar radio emission is believed to
be generated at frequencies of order of the local Lorentz-shifted
proper plasma frequency, $\omega\sim\omega_p\sqrt{\gamma}$, where
$\omega_p\equiv\sqrt{4\pi Ne^2/m}$ and $N$ is the number density
of the plasma particles. Therefore in the radio emission region
and its close vicinity the scattering is a collective plasma
process. The induced scatterings in the plasma are suggested as an
important ingredient of the pulsar radio emission mechanism
\citep[see, e.g.,][]{l79,l92,l93,lyub96,l98} and as a significant
propagation effect of the generated radiation
\citep{gk93,l98,lm06}. However, as the plasma number density
rapidly decreases with distance from the neutron star, $N\propto
R^{-3}$, far enough from the emission region
$\omega\gg\omega_p\sqrt{\gamma}$ and the plasma effects on the
scattering become negligible. Thus, the scattering in the
resonance region can be considered as a single-particle process.
Besides that, in this region the wave dispersion can also be
ignored, so that the incident radiation presents transverse
electromagnetic waves.

A general formalism of the scattering by gyrating electrons in the
magnetoactive plasma has been developed in \citet{ms72}. In the
present paper, we introduce some corrections and simplifications
to that treatment and obtain the scattering cross-section in the
form suitable for the concrete applications. The plan of the paper
is as follows. In Sect.~2 we derive the scattering cross-section
for a gyrating electron and obtain its low-frequency
approximation. The scattering of pulsar radio emission by
relativistic gyrating particles is studied in Sect.~3. We
investigate the validity of the formalism in application to
pulsars in Sect.~3.1, examine the radio intensity suppression in
Sect.~3.2 and consider the scattered power in Sect.~3.3. The
results of the paper are discussed and summarized in Sect.~4. The
induced scattering off the gyrating electrons will be considered
in a separate paper.

\section{Scattering cross-section}
\subsection{General equations}
Let us consider the scattering of transverse electromagnetic waves
by a gyrating electron. For the sake of simplification we take
that the component of the electron velocity along the external
magnetic field is zero. In the classical formulation of the
problem, the incident wave fields perturb the motion of the
electron, and it emits the secondary waves, which are interpreted
as a scattered radiation. We proceed from the formalism developed
in \citet{ll88} for the scattering by a free electron. The
scattering cross-section is defined as the ratio of the average
intensity of the waves scattered into the elementary solid angle
${\rm d}O^\prime$ to the energy flux density of the incident
radiation, and can be presented in the form
\begin{equation}
\frac{{\rm d}\sigma}{{\rm
d}O^\prime}=\frac{\lim\limits_{T\to\infty}\frac{R_0^2}{T}\int\limits_{-\infty}^\infty
\vert {\bmath k}^\prime\times {\bmath A}_{\omega^\prime}\vert^2
{\rm d}\omega^\prime /2\pi}{\vert {\bmath E}_\omega\vert^2},
\end{equation}
where ${\bmath k}^\prime$ and $\omega^\prime$ are the wavevector
and frequency of the scattered waves, ${\bmath A}_{\omega^\prime}$
and ${\bmath E}_\omega$ are the Fourier-components of the vector
potential of the scattered waves and the electric field of the
incident waves, and $R_0$ is the distance to an observer. The
linearized vector potential of the scattered waves reads
\begin{equation}
{\bmath A}_{\omega^\prime}=\frac{{\rm e}^{{\rm i} k^\prime
R_0}}{cR_0}\int\limits_{-\infty}^\infty e[{\bmath v}_1(t)-{\rm
i}({\bmath k}^\prime\cdot {\bmath r}_1(t)){\bmath v}_0(t)]{\rm
e}^{{\rm i}\omega^\prime t-{\rm i}{\bmath k}^\prime\cdot {\bmath
r}_0(t)}{\rm d}t.
\end{equation}
Here ${\bmath v}_0$ and ${\bmath r}_0$ are the velocity and
coordinate of the unperturbed circular motion of the electron,
${\bmath v_1}$ and ${\bmath r}_1$ are the first order
perturbations of these quantities. In the coordinate system with
the z-axis along the external magnetic field,
\begin{equation}
{\bmath v}_0=v_0(\cos \Omega t,-\sin\Omega t,0),\quad {\bmath
r}_0=\frac{v_0}{\Omega}(\sin\Omega t,\cos\Omega t,0),
\end{equation}
where $\Omega\equiv eB_0/\gamma_0mc$ and $\gamma_0\equiv
(1-v_0^2/c^2)^{-1/2}$. The perturbed motion of the particle in the
fields of the incident wave, ${\bmath E}_1$ and ${\bmath B}_1$, is
described by the linearized equation of motion
\begin{equation}
m\gamma_0\frac{\rm d}{{\rm d}t}\left ({\bmath
v}_1+\gamma_0^2\frac{{\bmath v}_0\cdot{\bmath v}_1}{c^2}{\bmath
v}_0\right ) =e{\bmath E}_1+e\frac{{\bmath v}_0\times {\bmath
B_1}}{c}+e\frac{{\bmath v}_1\times {\bmath B}_0}{c}.
\end{equation}
Hereafter the subscripts of the perturbed quantities will be
omitted. It is convenient to project equation (4) onto the axes,
one of which is along ${\bmath v}_0$ and another one along
${\bmath B}_0$:
\[{\bmath e}_1=(\cos\Omega t,-\sin\Omega t,0),\quad
{\bmath e}_2=(\sin\Omega t,\cos\Omega t,0),\quad {\bmath
e}_3=(0,0,1).\] Taking into account that ${\rm d}{\bmath e}_1/{\rm
d}t=-\Omega{\bmath e}_2$ and ${\rm d}{\bmath e}_2/{\rm
d}t=\Omega{\bmath e}_1$, one can write
\[ \gamma_0^2\frac{\rm d}{{\rm d}t}({\bmath v}\cdot {\bmath
e}_1)={\bmath F}\cdot {\bmath e}_1, \]
\[
\frac{\rm d}{{\rm d}t}({\bmath v}\cdot {\bmath
e}_2)-\Omega\beta_0^2\gamma_0^2({\bmath v}\cdot {\bmath
e}_1)={\bmath F}\cdot {\bmath e}_2, \]
\begin{equation}
\frac{\rm d}{{\rm d}t}({\bmath v}\cdot {\bmath e}_3)={\bmath
F}\cdot {\bmath e}_3,
\end{equation}
where ${\bmath F}\equiv e({\bmath E}+{\bmath v}_0\times {\bmath
B}/c)/m\gamma_0$ and $\beta_0\equiv v_0/c$. Then we obtain the
following solution:
\[ v_x=f_1\cos\Omega t+(f_2+g)\sin\Omega t,\]
\[ v_y=-f_1\sin\Omega
t+(f_2+g)\cos\Omega t,
\]
\begin{equation} v_z=f_3,
\end{equation}
where
\begin{equation}
f_1=\frac{1}{\gamma_0^2}\int\limits^t{\bmath F}\cdot {\bmath
e}_1(t^\prime){\rm d}t^\prime,\quad f_{2,3}=\int\limits^t{\bmath
F}\cdot {\bmath e}_{2,3}(t^\prime){\rm d}t^\prime,\quad
g=\Omega\beta_0^2\int\limits^t{\rm
d}t^\prime\int\limits^{t^\prime}{\bmath F}\cdot {\bmath
e}_1(t^{\prime\prime}){\rm d}t^{\prime\prime},
\end{equation}
and the perturbed coordinate is given by ${\bmath r}=\int^t{\bmath
v}(t^\prime){\rm d}t^\prime$. Note that our equation of motion and
its solution differ substantially from those given in \citet{ms72}
(cf. equations (51)-(54) therein). Firstly, in that paper, one of
the terms of the linearized equation of motion is missing (namely,
the last term in equation (4) above). Secondly, the authors have
not taken into account that ${\bmath e}_i(t){\rm d}{\bmath p}/{\rm
d}t\neq {\rm d} ({\bmath e}_i\cdot {\bmath p})/{\rm d}t$.
Consequently, the second term on the left-hand side of the second
of equations (5) and the term $g$ in equation (6) are absent in
their treatment.

To proceed further we specify the characteristics of the incident
and scattered waves. The wavevectors of the incident and scattered
radiation can be written as
\begin{equation}
{\bmath
k}=k(\sin\theta\cos\phi,\sin\theta\sin\phi,\cos\theta),\quad
{\bmath k}^\prime=k^\prime(\sin\theta^\prime\cos\phi^\prime,
\sin\theta^\prime\sin\phi^\prime,\cos\theta^\prime),
\end{equation}
where $\theta,\theta^\prime$ and $\phi,\phi^\prime$ are the polar
and azimuthal angles in the spherical coordinate system with the
polar axis along the external magnetic field. Since the plasma
effects are neglected, the radiation presents transverse
electromagnetic waves polarized either in the plane of the
wavevector and the external magnetic field or perpendicularly to
this plane. Then the electric fields of the waves are directed as
\[{\bmath
e}_A=(\cos\theta\cos\phi,\cos\theta\sin\phi,-\sin\theta),\quad
{\bmath e}_B=(\sin\phi,-\cos\phi,0),  \]
\begin{equation}
{\bmath e}_{A^\prime}=(\cos\theta^\prime\cos\phi^\prime,
\cos\theta^\prime\sin\phi^\prime,-\sin\theta^\prime),\quad {\bmath
e}_{B^\prime}=(\sin\phi^\prime,-\cos\phi^\prime,0),
\end{equation}
and ${\bmath B}={\bmath k}\times {\bmath E}$.

To find the vector potential (2) we use the Fourier-representation
of the monochromatic field, ${\bmath E}\propto {\bmath e}^{-{\rm
i}\omega t+{\rm i}{\bmath k}\cdot {\bmath r}_0(t)}$, in equation
(7), and perform the expansion in Bessel functions
\begin{equation}
{\rm e}^{{\rm i}{\bmath k}\cdot {\bmath r}_0}={\rm e}^{{\rm
i}(\omega/\Omega)\beta_0\sin\theta\sin(\Omega t+\phi)}
=\sum_{n=-\infty}^\infty J_n\left
(\frac{\omega\beta_0\sin\theta}{\Omega}\right ){\rm e}^{{\rm
i}n\Omega t}{\rm e}^{{\rm i}n\phi}
\end{equation}
and the analogous expansion for ${\rm e}^{-{\rm i}{\bmath
k}^\prime\cdot {\bmath r}_0(t)}$. With the commutation relations
$(J_{n+1}+J_{n-1})/2=(n/z)J_n$ and
$(J_{n-1}-J_{n+1})/2=J_n^\prime$, where $z$ is the argument of the
Bessel function and the prime denotes the derivative with respect
to $z$, one can obtain the useful relations
\[\cos(\Omega t+\phi){\rm e^{{\rm i}{\bmath k}\cdot{\bmath
r}_0}}=\sum_{n=-\infty}^\infty\frac{n}{z}J_n(z){\rm e}^{{\rm
i}n(\Omega t+\phi)},\quad \sin(\Omega t+\phi){\rm e^{{\rm
i}{\bmath k}\cdot{\bmath r}_0}}=-{\rm i }\sum_{n=-\infty}^\infty
J_n^\prime(z){\rm e}^{{\rm i}n(\Omega t+\phi)},\]
\[ \cos^2(\Omega
t+\phi^\prime){\rm e^{-{\rm i}{\bmath k}^\prime\cdot{\bmath
r}_0}}=\sum_{l=-\infty}^\infty\left
[\frac{l^2}{z^{\prime^2}}J_l(z^\prime)-\frac{J_l^\prime(z^\prime)}{z^\prime}\right
]{\rm e}^{-{\rm i}l(\Omega t+\phi^\prime)},\]
\[\sin^2(\Omega
t+\phi^\prime){\rm e^{-{\rm i}{\bmath k}^\prime\cdot{\bmath
r}_0}}=\sum_{l=-\infty}^\infty\left [\left (1-
\frac{l^2}{z^{\prime^2}}\right )
J_l(z^\prime)+\frac{J_l^\prime(z^\prime)}{z^\prime}\right ]{\rm
e}^{-{\rm i}l(\Omega t+\phi^\prime)},\] \[ \cos(\Omega
t+\phi^\prime)\sin(\Omega t+\phi^\prime){\rm e^{-{\rm i}{\bmath
k}^\prime\cdot{\bmath r}_0}}=-{\rm i}\sum_{l=-\infty}^\infty\left
[\frac{l}{z^{\prime^2}}J_l(z^\prime)-\frac{l}{z^\prime}J_l^\prime(z^\prime)\right
]{\rm e}^{-{\rm i}l(\Omega t+\phi^\prime)}, \] where
$z=\omega\beta_0\sin\theta/\Omega$ and
$z^\prime=\omega^\prime\beta_0\sin\theta^\prime/\Omega$.
Substituting this into equation (2) leads to the integral $\int
{\rm e}^{{\rm i}(\omega^\prime-\omega+n\Omega-l\Omega)t}{\rm d}t$,
which yields the delta-function. Then, taking into account that
$[\delta(\omega)]^2=(T/2\pi)\delta(\omega)$ and using this in
equation (1), one can find the scattering cross-section.

We are interested in the four cross-sections corresponding to the
cases when the incident and scattered waves have one of the two
polarizations given by equation (9). The incident polarization
enters ${\bmath F}$, whereas the scattered polarization can be
included by projecting the magnetic field of the scattered
radiation, ${\bmath B}_{\omega^\prime}={\rm i}{\bmath
k}^\prime\times {\bmath A}_{\omega^\prime}$, onto the magnetic
field directions in the normal waves, ${\bmath
k^\prime}\times{\bmath e}_{A^\prime,B^\prime}$. Then the term
$\vert{\bmath k}^\prime\times{\bmath A}_{\omega^\prime}\vert^2$ is
reduced to $({\bmath A}_{\omega^\prime}\cdot {\bmath
e}_{A^\prime,B^\prime})^2\omega^{\prime^2}/c^2$. Routine
calculations lead to the following cross-sections:
\begin{equation}
\frac{{\rm d}\sigma^{ij}}{{\rm
d}O^\prime}=\frac{r_e^2}{\gamma_0^2}\sum_{\nu=-\infty}^\infty\omega^{\prime^4}
\left\vert\sum_{n,l=-\infty \atop
l-n=\nu}^\infty\frac{a_{nl}^{ij}}{(n\Omega-\omega)^2}{\rm e}^{{\rm
i}(n\phi-l\phi^\prime)}\right\vert^2,
\end{equation}
where $\omega^\prime\equiv\omega+(l-n)\Omega$, $r_e$ is the
classical electron radius, the superscripts $i,j$ denote the
initial and final polarization states of the waves, and
\[a^{AA^\prime}={\rm
i}\cos\theta\cos\theta^\prime\frac{[(n\Omega-\omega)J_n^\prime
+\Omega (n/z)J_n ][(n\Omega-\omega)J_l^\prime
+\Omega(l/z^\prime)J_l ]}{(n\Omega-\omega)^2-\Omega^2}\]
\[ +{\rm
i}J_nJ_l[(\beta_0n/z-\sin\theta
)(\beta_0l/z^\prime-\sin\theta^\prime)+\cos\theta\cos\theta^\prime
nl/zz^\prime\gamma_0^2],\]
\[ a^{AB^\prime}=\cos\theta\frac{[(n\Omega-\omega)J_n^\prime +\Omega
(n/z)J_n ][(n\Omega-\omega)(\beta_0\sin\theta^\prime
-l/z^\prime)J_l -\Omega J_l^\prime
]}{(n\Omega-\omega)^2-\Omega^2}\]
\[-J_nJ_l^\prime[n\cos\theta/z\gamma_0^2+\beta_0\cos\theta^\prime(\beta_0n/z-\sin\theta)],\]
\[ a^{BA^\prime}=\cos\theta^\prime\frac{[\Omega J_n^\prime
-(n\Omega-\omega)(\beta_0\sin\theta-n/z)J_n
][(n\Omega-\omega)J_l^\prime +\Omega
(l/z^\prime)J_l]}{(n\Omega-\omega)^2-\Omega^2}\]
\[ +J_n^\prime
J_l[l\cos\theta^\prime/z^\prime\gamma_0^2
+\beta_0\cos\theta(\beta_0l/z^\prime-\sin\theta^\prime)],\]
\[ a^{BB^\prime}={\rm
i}\frac{[(n\Omega-\omega)(\beta_0\sin\theta-n/z)J_n -\Omega
J_n^\prime ][(n\Omega-\omega)(\beta_0\sin\theta^\prime
-l/z^\prime)J_l -\Omega J_l^\prime
]}{(n\Omega-\omega)^2-\Omega^2}\]
\begin{equation}
+{\rm
i}J_n^\prime J_l^\prime
(1+\beta_0^2\gamma_0^2\cos\theta\cos\theta^\prime)/\gamma_0^2.
\end{equation}
Here $J_n=J_n(z)$ and $J_l=J_l(z^\prime)$. Equations (11)-(12)
give the scattering cross-sections in case of a gyrating electron.
In astrophysical applications, the particles generally perform
helical motion, and the corresponding cross-sections can be
obtained from equations (11)-(12) by means of relativistic
transformations. In case of relativistic longitudinal motion of
the electron, $\gamma_\Vert=(1-\beta_\Vert^2)^{-1/2}\gg 1$, it is
convenient to involve the cross-section $\Sigma$ defined as the
ratio of the number of the scattered photons to the flux density
of the photons flying against the electron. This quantity is a
relativistic invariant and is related to the cross-section
$\sigma$ as $(\omega/\omega^\prime){\rm d}\sigma/{\rm
d}O^\prime=(1-\beta_\Vert\cos\theta){\rm d}\Sigma/{\rm
d}O^\prime$. Then making use of the transformations
$\omega_c=\omega\gamma_\Vert\eta$,
$\omega_c^\prime=\omega^\prime\gamma_\Vert\eta^\prime$, and ${\rm
d }O_c^\prime={\rm d}O^\prime/\gamma_\Vert^2\eta^{\prime^2}$
(where $\eta\equiv 1-\beta_\Vert\cos\theta$, $\eta^\prime\equiv
1-\beta_\Vert\cos\theta^\prime$, and the quantities of the guiding
centre frame are denoted by the subscript 'c'), one can obtain
that
\begin{equation}
\frac{{\rm d}\sigma}{{\rm d}O^\prime}=\left (\frac{{\rm
d}\sigma}{{\rm d}O^\prime}\right
)_c\frac{\eta^2}{\gamma_\Vert^2\eta^{\prime^3}},
\end{equation}
where the quantities entering $({\rm d}\sigma/{\rm d}O^\prime)_c$
should be expressed via the quantities of the laboratory frame.

It is worthy to examine the symmetry properties of the
cross-sections (11)-(12). Keeping in mind that $n\Omega
-\omega\equiv l\Omega-\omega^\prime$, one can see that $\vert
a_{nl}^{ij}\vert$ are symmetrical with respect to the simultaneous
change $\omega{\iff}\omega^\prime$, $n{\iff}l$, and $i{\iff} j$,
and hence the cross-sections can be written in the form ${\rm
d}\sigma^{ij}/{\rm d
}O^\prime=\sum_{\nu}s_\nu^{ij}\omega^{\prime^4}$, where
$s_\nu^{ij}$ are symmetrical in  the above mentioned sense. This
corresponds to the symmetry of each harmonic of the scattering
probability with respect to the initial and final photon states.
Indeed, the power supplied by the scattering electron can be
presented as
\begin{equation}
P=\int\frac{{\rm d}\sigma}{{\rm d}O^\prime}{\rm d}O^\prime {\rm
d}\omega^\prime c\hbar\omega N({\bmath k})\frac{{\rm d}^3{\bmath
k}}{(2\pi)^3} =\int\hbar\omega^\prime w N({\bmath k})\frac{{\rm
d}^3{\bmath k}}{(2\pi)^3}\frac{{\rm d}^3{\bmath
k^\prime}}{(2\pi)^3},
\end{equation}
where $N({\bmath k})$ is the occupation number of the incident
photons and $w$ is the scattering probability. With equations
(13)-(14) it is obvious that $w_\nu\propto
s_\nu\omega\omega^\prime\eta\eta^\prime$.

\subsection{Useful approximations}
The general form of the scattering cross-section given by
equations (11)-(12) is so complicated that it can hardly be
involved directly in concrete applications. To analyze the basic
features of the scattering off a gyrating electron we turn to
reasonable approximations. (Note that the approximate
cross-sections obtained below refer to the guiding-centre frame
and it is necessary to apply the Lorentz transformation (13) in
order to use them in any realistic calculations.) First of all, we
consider the limiting case when $\beta_0\to 0$. As $z,z^\prime \to
0$, one can use the approximation of the Bessel function at small
arguments,
\begin{equation}
J_n(\zeta)\approx\frac{(\zeta/2)^n}{n!},\quad \zeta\ll 1, n\geq 0,
\end{equation}
taking into account that $J_{-n}(\zeta)=(-1)^nJ_n(\zeta)$. Then
only the zeroth harmonic, $\nu=0$ ($\omega^\prime=\omega$), yields
a non-zero contribution to the cross-sections, and they are
reduced to the form
\[ \frac{{\rm d}\sigma^{AA^\prime}}{{\rm
d}O^\prime}=r_e^2\left\vert\sin\theta\sin\theta^\prime+\cos\theta\cos\theta^\prime\frac{{\rm
i }\Omega\omega\sin\Delta\phi-\omega^2\cos\Delta\phi}
{\Omega^2-\omega^2}\right\vert^2,\]
\[ \frac{{\rm
d}\sigma^{AB^\prime}}{{\rm
d}O^\prime}=\frac{r_e^2\omega^4\cos^2\theta}
{(\Omega^2-\omega^2)^2}\left
[\frac{\Omega^2}{\omega^2}\cos^2\Delta\phi+\sin^2\Delta\phi\right
],\]
\[ \frac{{\rm d}\sigma^{BA^\prime}}{{\rm
d}O^\prime}=\frac{r_e^2\omega^4\cos^2\theta^\prime}
{(\Omega^2-\omega^2)^2}\left
[\frac{\Omega^2}{\omega^2}\cos^2\Delta\phi+\sin^2\Delta\phi\right
],\]
\begin{equation}\frac{{\rm
d}\sigma^{BB^\prime}}{{\rm d}O^\prime}=\frac{r_e^2\omega^4}
{(\Omega^2-\omega^2)^2}\left
[\frac{\Omega^2}{\omega^2}\sin^2\Delta\phi+\cos^2\Delta\phi\right
],
\end{equation}
where $\Delta\phi\equiv\phi-\phi^\prime$. These equations coincide
with the scattering cross-sections for the electron at rest
\citep{c71}.

Below we dwell on the low-frequency approximation of equation
(12), $\omega\ll\Omega$, given arbitrary gyration velocities of
the electron. (Note that our assumption does imply small
frequencies rather than large gyrofrequencies, since in the latter
case the electrons rapidly lose their gyration energies via
synchrotron emission.) Then $z$ is still a small quantity, while
$z^\prime=\omega^\prime\beta_0\sin\theta^\prime/ \Omega$ is small
only at $\nu=0$; at $\nu\neq 0$
$z^\prime\approx\nu\beta_0\sin\theta^\prime$. The zeroth-harmonic
cross-sections read
\[ \frac{{\rm d}\sigma_0^{AA^\prime}}{{\rm
d}O^\prime}=\frac{r_e^2\sin^2\theta\sin^2\theta^\prime}{\gamma_0^2},
\]
\[ \frac{{\rm d}\sigma_0^{AB^\prime}}{{\rm
d}O^\prime}=\frac{r_e^2\omega^2}{\gamma_0^2\Omega^2}\left
[\cos\theta\cos\Delta\phi -\frac{\beta_0^2\sin\theta
\sin\theta^\prime\cos\theta^\prime}{2}\right ]^2, \]
\[ \frac{{\rm d}\sigma_0^{BA^\prime}}{{\rm
d}O^\prime}=\frac{r_e^2\omega^2}{\gamma_0^2\Omega^2}\left
[\cos\theta^\prime\cos\Delta\phi -\frac{\beta_0^2\sin\theta
\sin\theta^\prime\cos\theta}{2}\right ]^2, \]
\begin{equation}
\frac{{\rm d}\sigma_0^{BB^\prime}}{{\rm
d}O^\prime}=\frac{r_e^2\omega^2}{\gamma_0^2\Omega^2}\sin^2\Delta\phi.
\end{equation}
Comparison of equation (17) with equation (16) at
$\omega/\Omega\ll 1$ shows the following. For the scattering
channel $B\to B^\prime$, the cross-sections are exactly the same,
i.e. the scattering by the gyrating electron without a change in
frequency nothing differs from that by the electron at rest. For
the scattering channels with a change in polarization state, $A\to
B^\prime$ and $B\to A^\prime$, the cross-sections (16) and (17)
are somewhat different, though generally remain of the same order
of magnitude. As for the $A\to A^\prime$ channel, the scattering
by a gyrating electron is a factor of $\gamma_0^{-2}$ weaker.
Nevertheless, it may still dominate that in the other channels.

The cross-section components at the non-zero harmonics can be
reduced to

\[ \frac{{\rm d}\sigma_\nu^{AA^\prime}}{{\rm
d}O^\prime}=\frac{r_e^2\Omega^4\nu^4J^2_\nu(\nu\beta_0\sin\theta^\prime)}
{\gamma_0^2\omega^4}\frac{\sin^2\theta\cos^4\theta^\prime}{\sin^2\theta^\prime},
\]
\[ \frac{{\rm d}\sigma_\nu^{AB^\prime}}{{\rm
d}O^\prime}=\frac{r_e^2\Omega^4\nu^4J^{\prime^2}_\nu(\nu\beta_0\sin\theta^\prime)}
{\gamma_0^2\omega^4}\beta_0^2\sin^2\theta\cos^2\theta^\prime, \]
\[ \frac{{\rm d}\sigma_\nu^{BA^\prime}}{{\rm
d}O^\prime}=\frac{r_e^2\Omega^2\nu^4J^2_\nu(\nu\beta_0\sin\theta^\prime)}
{\gamma_0^2\omega^2}\frac{\cos^2\theta^\prime}{\sin^2\theta^\prime}
\{\sin\theta [1-\beta_0^2(1-\cos\theta\cos\theta^\prime)/2
]+\sin\theta^\prime\cos\Delta\phi\}^2, \]
\begin{equation}
\frac{{\rm d}\sigma_\nu^{BB^\prime}}{{\rm
d}O^\prime}=\frac{r_e^2\Omega^2\nu^4J^{\prime^2}_\nu(\nu\beta_0\sin\theta^\prime)\beta_0^2}
{\gamma_0^2\omega^2}\frac{\cos^2\theta^\prime}{\sin^2\theta^\prime}
\{\sin\theta [1-\beta_0^2(1-\cos\theta\cos\theta^\prime)/2
]+\sin\theta^\prime\cos\Delta\phi\}^2,
\end{equation}
and the total cross-sections are given by
\begin{equation}
\frac{{\rm d}\sigma^{ij}}{{\rm d}O^\prime}=\frac{{\rm
d}\sigma_0^{ij}}{{\rm d}O^\prime}+2\sum_{\nu=1}^\infty\frac{{\rm
d}\sigma_\nu^{ij}}{{\rm d}O^\prime}.
\end{equation}
Note that the dependence of the cross-section components (18) on
$\nu$ resembles that of the harmonics of the synchrotron power in
the Schott formula, except for the factor $\nu^4$ instead of
$\nu^2$. Based on an analysis similar to that in the classical
theory of synchrotron emission, we conclude that the scattering
cross-sections peak at high harmonics, $\nu\sim\gamma_0^3$, and
the low-frequency radiation, $\omega\ll\Omega$, scattered by the
relativistically gyrating electron concentrates in the same
frequency range as the synchrotron emission of the electron.

To obtain the total scattering cross-sections we make use of the
summation formulas derived in Appendix,
\[
\sum_{\nu=1}^\infty\nu^4J_\nu^2(\nu
x)=\frac{x^2(64+592x^2+472x^4+27x^6)}{256(1-x^2)^{13/2}},\]
\begin{equation}
\sum_{\nu=1}^\infty\nu^4J_\nu^{\prime^2}(\nu
x)=\frac{64+624x^2+632x^4+45x^6}{256(1-x^2)^{11/2}},
\end{equation}
and perform routine integration over the solid angle. The final
results read
\[ \sigma^{AA^\prime}=\frac{4\pi
r_e^2\Omega^4\gamma_0^6\beta_0^2\sin^2\theta}{\omega^4}\left
(\frac{1}{10}+\frac{\beta_0^2}{20}-\frac{\beta_0^4}{30}+\frac{\beta_0^6}{120}\right
),\]
\[ \sigma^{AB^\prime}=\frac{4\pi
r_e^2\Omega^4\gamma_0^6\beta_0^2\sin^2\theta}{\omega^4}\left
(\frac{1}{6}+\frac{7\beta_0^2}{20}+\frac{\beta_0^4}{30}-\frac{\beta_0^6}{120}\right
), \]
\[ \sigma^{BA^\prime}=\frac{4\pi
r_e^2\Omega^2\gamma_0^8\beta_0^2}{\omega^2}\left\{\left
[\frac{1}{2}+\sin^2\theta\left (1-\frac{\beta_0^2}{2}\right
)^2\right ]\left
(\frac{1}{6}+\frac{13\beta_0^2}{60}-\frac{\beta_0^4}{12}+
\frac{\beta_0^6}{24}-\frac{\beta_0^8}{120}\right )\right.\]
\[
\left.+\frac{1}{\gamma_0^2}\left
(\frac{\beta_0^4\cos^2\theta\sin^2\theta}{4}-\frac{1}{2}\right
)\left
(\frac{1}{10}+\frac{\beta_0^2}{20}-\frac{\beta_0^4}{30}+\frac{\beta_0^6}{120}\right
) \right\}, \]
\[ \sigma^{BB^\prime}=\frac{4\pi
r_e^2\Omega^2\gamma_0^8\beta_0^2}{\omega^2}\left\{\left
[\frac{1}{2}+\sin^2\theta\left (1-\frac{\beta_0^2}{2}\right
)^2\right ]\left
(\frac{1}{2}+\frac{31\beta_0^2}{12}+\frac{77\beta_0^4}{60}-
\frac{\beta_0^6}{24}+\frac{\beta_0^8}{120}\right )\right.\]
\begin{equation}
\left.+\frac{1}{\gamma_0^2}\left
(\frac{\beta_0^4\cos^2\theta\sin^2\theta}{4}-\frac{1}{2}\right
)\left
(\frac{1}{6}+\frac{7\beta_0^2}{20}+\frac{\beta_0^4}{30}-\frac{\beta_0^6}{120}\right
) \right\}.
\end{equation}
The cross-sections include the large parameters,
$\Omega^4/\omega^4$ and $\Omega^2/\omega^2$, and strongly exceed
those for the electron at rest, but it should be kept in mind that
they cannot increase unrestrictedly with $\Omega$, since strong
enough magnetic field precludes the relativistic gyration of the
electron. As can be seen from equation (21), the cross-sections
for the same incident polarizations are of the same order of
magnitude and in case of ultrarelativistic gyration,
$\beta_0\approx 1$, are related as
$\sigma^{AA^\prime}/\sigma^{AB^\prime}=3/13$ and
$\sigma^{BA^\prime}/\sigma^{BB^\prime}=1/13$, whereas
$(\sigma^{AA^\prime}+\sigma^{AB^\prime})/(\sigma^{BA^\prime}+\sigma^{BB^\prime})
=(16/9)\Omega^2/\omega^2\gamma_0^2$. The latter quantity may be
both small and large, so that any of the two polarizations may be
scattered predominantly. Stronger scattering is favored by larger
gyration energies and larger $\Omega/\omega$, and the scattering
of B-polarization is more significantly affected by the former
quantity, whereas the scattering of A-polarization by the latter
one.

\section{Application to pulsars}
The scattering of low-frequency radiation by relativistic gyrating
particles is believed to be the case in pulsars. It takes place in
the outer magnetosphere, in the region of cyclotron resonance for
the high-frequency radio waves. The particles of the secondary
plasma, which pass through the area covered by the pulsar radio
beam, participate in cyclotron absorption and rapidly acquire
relativistic rotational energies. One can expect that these
particles subsequently scatter the low-frequency radio waves,
which are still below the resonance, in the regime considered in
Sect.~2.2. Below we examine the validity of the scattering
cross-section obtained above in application to pulsars, give the
quantitative description of the consequences of the scattering
process and compare them with the consequences of resonant
absorption.

\subsection{Validity of the formalism}
We start from examining the validity of the technique developed in
Sect.~2.1. The linearization of the equation of the particle
motion is appropriate on condition that
\begin{equation}
v_1\ll v_0\sim c \qquad {\rm and}\qquad r_1\ll r_0\sim c/\Omega.
\end{equation}
In case of the B-polarized incident waves, when the incident
electric field is perpendicular to the external magnetic field,
from equation (4) one can obtain the following estimate: $v_1\max
(m\gamma\omega, eB_0/c)\sim eB_1$, which reduces to
\begin{equation}
\frac{v_1}{c}\sim\left\{\begin{array}{ll}B_1/B_0,&\omega/\Omega\ll
1,\\ eB_1/\omega mc\gamma,&\omega/\Omega\gg 1.\end{array}\right.
\end{equation}
The estimate at $\omega/\Omega\gg 1$ is also true in the absence
of the external magnetic field and coincides with the parameter
$f$ of the theory of synchro-Compton emission
\citep{go71,r71,b72}. The restriction given by equation (22)
implies that $f\ll 1$, i.e. the linearization technique is
admissible only in the case when the secondary waves emitted by
the electron present a customary Compton-scattered radiation
rather than the synchrotron emission in the magnetic field of a
strong low-frequency incident wave.

Our consideration is concerned with the regime of customary
Compton scattering in a strong external magnetic field. As can be
seen from equation (23), this regime is realized on condition
$B_1/B_0\ll 1$, i.e. if the energy density of the incident
radiation is much less than the energy density of the external
field. In case of pulsars, it is reasonable to analyze this
condition directly in the laboratory frame, where the particles
have the longitudinal component of the velocity as well. In this
case, the general form of equation (4) remains the same, but the
particle velocity and Lorentz-factor, ${\bmath v}_0$ and
$\gamma_0$, should be replaced by the total velocity ${\bmath v}$,
which includes both the longitudinal and transverse components,
$\beta^2=v^2/c^2=\beta_\Vert^2+\beta_\perp^2$, and the
corresponding Lorentz-factor $\gamma=(1-\beta^2)^{-1/2}$. Then for
the quantities of the laboratory system we have: $v_1/c\sim
B_1/B_0$ at $2\pi\nu_l\eta\gamma/\omega_H$, where $\nu_l$ is the
frequency of the incident waves and $\omega_H\equiv eB_0/mc$. The
magnetic field of a pulsar has dipolar structure and its strength
can be estimated as
\[ B_0=10^6\frac{B_\star}{10^{12}\,{\rm
G}}\left (\frac{R}{10^8\,{\rm cm }}\right )^{-3}\,{\rm G},
\]
where $B_\star$ corresponds to the stellar surface, $R$ is the
distance from the star, and it is taken that the neutron star
radius is $10^6$ cm. Taking into account that $B_1^2/8\pi=L/Sc$,
where $L$ is the radio luminosity, $S$ is the area of the radio
beam cross-section, $S=\pi R^2\chi^2$, and $\chi$ is the
half-width of the beam, one can estimate the magnetic field
strength in the waves as
\[ B_1=5\chi^{-1}\left
(\frac{L}{10^{27}\,{\rm erg\, s^{-1}}}\right )^{1/2}\left
(\frac{R}{10^8\, {\rm cm}}\right )^{-1}\,{\rm G}.
\]
Hence,
\begin{equation}
\frac{B_1}{B_0}=5\times 10^{-6}\chi^{-1}\left
(\frac{L}{10^{27}\,{\rm erg\, s^{-1}}}\right
)^{1/2}\frac{10^{12}\,{\rm G}}{B_\star}\left (\frac{R}{10^8\, {\rm
cm}}\right )^2.
\end{equation}
As is pointed out above, the scattering of low-frequency waves off
the gyrating particles occurs in the region of cyclotron
absorption of high-frequency waves, where the particles can
acquire substantial transverse energies. Its location in the
magnetosphere can be estimated from the resonance condition for
the high-frequency waves, $\nu_h\gamma\theta^2/2=\omega_H/2\pi$
(here it is taken into account that $\gamma=\gamma_0\gamma_\Vert$,
which follows from the invariance of the transverse momentum,
$p_\perp=\beta_\perp\gamma mc=\beta_0\gamma_0 mc$, and assumed
that $1/\gamma\ll\theta <1$):
\begin{equation}
\frac{R}{10^8\,{\rm cm}}=2\theta^{-2/3}\left
(\frac{B_\star}{10^{12}\,{\rm
G}}\frac{10^2}{\gamma}\frac{10^{10}\,{\rm Hz}}{\nu_h}\right
)^{1/3}.
\end{equation}
Using equations (24)-(25) and taking into account that $\chi\sim
n\cdot 0.1$, one can conclude that for any conceivable values of
the parameters $B_1/B_0$ is less than unity and, correspondingly,
the energy density of the incident radiation is much less than the
energy density of the external magnetic field.

Assuming that $r_1\sim v_1/\omega$, one can find that
\begin{equation}
\frac{r_1}{r_0}\sim\frac{B_1}{B_0}\frac{\Omega}{\omega}, \qquad
\frac{\omega}{\Omega}\ll 1.
\end{equation}
Thus, $r_1/r_0\ll 1$ appears a more restrictive condition than
$v_1/v_0\ll 1$ because of the large parameter $\Omega/\omega$.
However, it should be kept in mind that in our case the cyclotron
frequency should be small enough to provide the resonance of the
high-frequency radio waves, so that
$\Omega/\omega\sim\nu_h\theta^2/2\nu_l$. Since pulsar radio
emission spans the frequency range $n\cdot 10^7-n\cdot 10^{10}$
Hz, $\nu_h/\nu_l\sim 10^2-10^3$, and the condition $r_1/r_0\ll 1$
is also fulfilled.

The linearization of the vector potential given by equation (2) is
valid under the condition
\begin{equation}
{\bmath k}^\prime\cdot{\bmath
r}_1\sim\frac{\omega^\prime}{\omega}\frac{B_1}{B_0}\ll 1, \qquad
\frac{\omega}{\Omega}\ll 1,
\end{equation}
which is still more restrictive, since the waves are predominantly
scattered into the frequency range
$\omega^\prime\sim\gamma_0^3\Omega$.

Given that the incident waves have A-polarization, the electric
field and the perturbed velocity are almost aligned with $B_0$,
and the restrictions take the form
\[\frac{v_1}{v_0}\sim\frac{B_1}{B_0}\frac{\Omega}{\omega},\]
\[\frac{r_1}{r_0}\sim\frac{B_1}{B_0}\frac{\Omega^2}{\omega^2},\]
\begin{equation}
k^\prime
r_1\sim\frac{B_1}{B_0}\frac{\Omega^2}{\omega^2}\gamma_0^3.
\end{equation}
Based on the above consideration one can conclude that our
treatment is valid for $\gamma_0<10$. Such rotational energies are
indeed typical of the particles of the secondary plasma in the
magnetosphere of a pulsar \citep[see, e.g.,][]{p02,p03}.

In the guiding centre frame, the scattered power can be estimated
as $P^j=L\sigma/S$, where $j$ denotes the polarization state of
the incident waves. It is reasonable to compare $P^j$ with the
synchrotron power of the electron, $P_{\rm syn}\sim
e^2\omega_H^2\gamma_0^2/c$. Making use of the cross-sections (21)
at $\beta_0\approx 1$ and taking into account that
$L/Sc=B_1^2/8\pi$, one can obtain
\[\frac{P^A}{P_{\rm
syn}}\sim\frac{B_1^2}{B_0^2}\frac{\Omega^4}{\omega^4}\gamma_0^4,\]
\begin{equation}
\frac{P^B}{P_{\rm
syn}}\sim\frac{B_1^2}{B_0^2}\frac{\Omega^2}{\omega^2}\gamma_0^6.
\end{equation}
Keeping in mind the restrictions (23), (26)-(28), one can see that
the scattered power is always less than the synchrotron power. As
$k^\prime r_1$ approaches unity, $P^B/P_{\rm syn}\sim 1$ and
$P^A/P_{\rm syn}\sim\gamma_0^{-2}$.

\subsection{Intensity transfer}
As is shown in Sect.~2.2, the low-frequency waves,
$\omega_c\ll\Omega$, are predominantly scattered into the range of
high harmonics of the gyrofrequency,
$\omega_c^\prime\sim\gamma_0^3\Omega$. In the laboratory frame
$\omega^\prime\gamma_\Vert\eta^\prime=\omega_H\gamma_0^2$, where
$\eta^\prime\sim 1/\gamma_\Vert^2$ because of relativistic beaming
effect. Taking into account that
$\omega_H/\gamma_0=2\pi\nu_h\gamma_\Vert\eta$, one can estimate
the characteristic frequency of the scattered radiation as
\begin{equation}
\nu^\prime\sim10^{17}\theta^2\frac{\nu_h}{10^{10}\,{\rm Hz}}\left
(\frac{\gamma_\Vert}{10^2}\right )^2\left
(\frac{\gamma_0}{10}\right )^3\,{\rm Hz}.
\end{equation}
It strongly depends on the particle rotational energy $\gamma_0$
and other parameters and is expected to fall into the optical or
soft X-ray range. Note that the scattering of the low-frequency
radiation into the high-energy band strongly dominates the inverse
process, since the radio emission of a pulsar is much more
intense.

The scattering depth can be written as
\begin{equation}
\Gamma^i=\int\Sigma\eta N_e{\rm
d}R=\int\frac{\omega\eta}{\omega^\prime\eta^\prime}\sigma_c^i\eta
N_e{\rm d}R,
\end{equation}
where $N_e$ is the number density of the scattering particles and
$\sigma_c^i$ stands for the scattering cross-sections (21)
expressed in terms of the quantities of the laboratory frame and
summed over the final polarization states:
\[ \sigma^A=32\pi
r_e^2\gamma_0^6(\nu_h/\nu_l)^4/3\theta^2\gamma_\Vert^2,\]
\begin{equation}
\sigma^B=28\pi r_e^2\gamma_0^8(\nu_h/\nu_l)^2/3.
\end{equation}
Here it is taken that $\beta_0\approx 1$, $1/\gamma_\Vert\ll\theta
<1$, $\Omega=2\pi\nu_h\gamma_\Vert\eta$, and
$\omega_c=2\pi\nu_l\gamma_\Vert\eta$. The number density of pulsar
plasma can be presented in terms of the Goldreich-Julian number
density, $N_e=\kappa B_0/2\pi R_Le$ (where $\kappa$ is the plasma
multiplicity factor and $R_L=5\cdot 10^9P$ cm is the light
cylinder radius), and estimated as
\begin{equation}
N_e=6.4\times 10^7\frac{\kappa}{10^3}\frac{B_\star}{10^{12}\,{\rm
G }}\frac{1\,{\rm s}}{P}\left (\frac{R}{10^8\,{\rm cm}}\right
)^{-3}\,{\rm cm}^{-3}.
\end{equation}
Substituting equations (32)-(33) into equation (31) and taking
into account that
$\omega\eta/\omega^\prime\eta^\prime\sim(\nu_l/\nu_h)\gamma_0^{-3}$,
we find
\[ \Gamma^A=10^{-6}\gamma_0^3\left (\frac{\nu_h/\nu_l}{10^2}\right
)^3\left (\frac{10^2}{\gamma_\Vert}\frac{10^8\,{\rm cm}}{R}\right
)^2 \frac{\kappa}{10^3}\frac{B_\star}{10^{12}\,{\rm
G}}\frac{1\,{\rm s}}{P},\]
\begin{equation}
\Gamma^B=10^{-6}\gamma_0^5\frac{\nu_h/\nu_l}{10^2}
\frac{\kappa}{10^3}\frac{B_\star}{10^{12}\,{\rm G }}\frac{1\,{\rm
s}}{P}\left (\frac{10^8\,{\rm cm}}{R}\right )^{2}\theta^2.
\end{equation}
One can see that the intensity suppression can be efficient only
marginally, at $\gamma_0\sim 10$ and $\nu_h/\nu_l\sim10^3$. Note
that the scattering depths for the two polarizations have distinct
dependencies on the parameters and may differ substantially.
Hence, if efficient, the scattering should affect polarization of
outgoing radiation, since the latter is an incoherent mixture of
the waves with the two polarization states.

It is interesting to compare the above scattering efficiencies
with the optical depth to resonant absorption. The latter quantity
is given by
\begin{equation}
\Gamma_c=2\frac{\kappa}{10^3}\left (\frac{B_\star}{10^{12}\,{\rm G
}}\frac{10^9\,{\rm Hz}}{\nu}\frac{10^2}{\gamma}\right )^{3/5}
\left (\frac{1\,{\rm s}}{P}\right )^{9/5}\sin^{4/5}\xi,
\end{equation}
where $\xi$ is the angle between the rotational and magnetic axes
of a pulsar \citep[see, e.g., equation (2.8) in ][]{lp98}. Given
that the plasma effects are ignored, the absorption depth is the
same for the waves of the two polarizations. As is obvious from
equation (35), resonant absorption can markedly affect the
intensity of pulsar radio emission, especially at low enough
frequencies. Note also that the scattering efficiencies depend on
the frequency stronger than the absorption depth, so that the
scattering can noticeably contribute to intensity suppression at
the lowest radio frequencies.

\subsection{Power of the scattered radiation}
For a single electron, the scattered power can be written as
\begin{equation}
P^j\approx\int\sigma_c^j\frac{\eta^2}{\eta^\prime}I_\omega(\omega,\theta,\phi)
{\rm d}\omega {\rm d}O,
\end{equation}
where $\eta^\prime\sim 1/\gamma_\Vert^2$,
$I_\omega(\omega,\theta,\phi)$ is the spectral intensity of the
incident radiation, $j$ is the final polarization state of the
scattered radiation, and $\sigma_c^j$ stands for the sum over the
two initial polarizations. The radio emission of a pulsar is
generated by the plasma and because of relativistic beaming is
concentrated into a narrow cone of the opening angle $\sim
1/\gamma_\Vert$. Therefore one can assume that the angular
distribution of the incident radiation is characterized by the
delta-function, which peaks at the angle $\theta$ with respect to
the external magnetic field ($1/\gamma_\Vert\ll\theta <1$) and at
an arbitrary azimuth. Pulsar spectra are generally described by
the power law,
\begin{equation}
I_\omega=I_{\omega_0}\left (\frac{\omega}{\omega_0}\right
)^{-\alpha},
\end{equation}
where the spectral index $\alpha$ ranges from 1 to 3 and
\begin{equation}
I_{\omega_0}\approx\frac{L}{2\pi\nu_0S},
\end{equation}
with $\nu_0$ corresponding to the low-frequency turnover in the
spectrum, $\nu_0\sim 10^8$ Hz.

Since the scattering cross-sections are also decreasing functions
of frequency, the main contribution to the integral in equation
(36) comes from the lowest frequencies, and we have
\begin{equation}
P^j\sim\frac{L}{S}\sigma_c^j(\nu_l)\gamma_\Vert^2\eta^2.
\end{equation}
In our case it is reasonable to take that $\nu_h/\nu_l/\gamma_0\gg
1$. Then the scattering of the A-polarization dominates and we
obtain \[ P^{A^\prime}=10^{-6}\gamma_0^6\left
(\frac{\nu_h/\nu_l}{10^2}\right )^4\frac{L}{10^{27}\,{\rm
erg\,s}^{-1}}\left (\frac{R}{10^8\,{\rm cm}}\right
)^{-2}\theta^2\chi^{-2}\,{\rm erg\,s}^{-1}, \]
\begin{equation}
P^{B^\prime}=\frac{13}{3}P^{A^\prime}.
\end{equation}
The total power provided by the system of the scattering
particles, $P\sim (P^{A^\prime}+P^{B^\prime})N_eSR$, is estimated
as
\begin{equation}
P=4\times 10^{26}\gamma_0^6\left (\frac{\nu_h/\nu_l}{10^2}\right
)^4\frac{L}{10^{27}\,{\rm erg\,s}^{-1}}\left (\frac{R}{10^8\,{\rm
cm}}\right )^{-2}\frac{\kappa}{10^3}\frac{B_\star}{10^{12}\,{\rm
G}}\frac{1\,{\rm s}}{P}\theta^2\,{\rm erg\,s}^{-1}.
\end{equation}
Although the scattered power does not exceed the synchrotron power
of the particles (see the end of Sect.~3.1), it may be large
enough to be observable. Thus, the scattering of low-frequency
radiation by gyrating particles may be an additional mechanism of
the pulsar high-energy emission.

As has been demonstrated in \citet{p02,p03}, the evolution of the
particle distribution function is mainly determined by the
resonant absorption, whereas the contribution of the spontaneous
synchrotron re-emission is typically insignificant. Hence, the
influence of the scattering on the particle momenta is all the
more weak. Note, however, that the induced scattering off the
gyrating particles, which is believed to be efficient because of
extremely high brightness temperatures of pulsar radiation, may
somewhat contribute to the increase of the particle rotational
energies; this point will be examined in detail elsewhere.

\section{Discussion and conclusions}
We have examined the magnetized scattering off a gyrating electron
and have particularly concentrated on the low-frequency case, when
the incident waves are well below the cyclotron resonance. The
electron gyration changes the character of the scattering
essentially: The scattered radiation presents a series of the
harmonics of the gyrofrequency, $\omega_c^\prime
=\omega_c+\nu\Omega$, $\nu=0,1,2\dots$. At $\nu=0$, the scattering
cross-sections for different polarization channels resemble those
in case of the electron at rest, but do not coincide exactly,
except for the channel $B\to B^\prime$. Furthermore, the
scattering cross-sections appear to peak at high harmonics,
$\nu\sim\gamma_0^3$, so that the scattered radiation concentrates
in the same range as the synchrotron emission of the electron,
$\omega_c^\prime\sim\gamma_0^3\Omega$. The total scattering
cross-section summed over the harmonics greatly exceeds the
magnetized cross-section for the electron at rest, and the
scattering process has distinct polarization signatures. In case
of the electron at rest, the scattering in the channel $A\to
A^\prime$ strongly dominates, whereas the cross-sections for the
other channels are of the same order, $\sim\omega_c^2/\Omega^2$
less, and differ from each other by geometrical factors. For the
relativistic gyrating electron, the cross-sections for the two
incident polarizations are related as
$\sigma^A/\sigma^B=16\Omega^2/9\omega_c^2\gamma_0^2$; this ratio
can be small and large, and
$\sigma^{AA^\prime}/\sigma^{AB^\prime}=3/13$,
$\sigma^{BA^\prime}/\sigma^{BB^\prime}=1/13$.

The low-frequency scattering off the particles performing
ultrarelativistic helical motion is directly applicable to
pulsars. Close to the neutron star surface, the magnetic field is
so strong that the particle rotational energies are almost
immediately lost via synchrotron emission. However, in the outer
magnetosphere, where the synchrotron losses are much less, the
particles can acquire substantial transverse energies as a result
of resonant absorption of the pulsar radio emission. The
scattering in the regime under consideration is believed to take
place at the bottom of the resonance region of radio waves, where
the high-energy waves meet the condition of cyclotron resonance,
whereas the low-frequency ones are still well below the resonance.

It is known that in pulsar case the spontaneous scattering by the
rectilinearly moving particles affects the radio intensities
negligibly. Although the scattering cross-section for the gyrating
particles is much larger, the scattering efficiency still remains
small over the radio frequency range, except for the lowest
frequencies. It should be noted that the resonant absorption can
noticeably suppress the radio intensities, especially at low
frequencies, and can account for the low-frequency turnovers in
pulsar spectra \citep{melr_a,melr_b}. Since the scattering
efficiencies for both polarization states are stronger functions
of frequency than the absorption depth, one can expect that the
scattering may contribute to intensity suppression beyond the
spectral turnover. Note that in contrast to the resonant
absorption the scattering affects the polarization of outgoing
radiation. The scattering signatures in pulsar radio emission are
yet to be studied observationally. At present, the range beyond
the low-frequency turnover is accessible only for a few radio
telescopes \citep*{konov1,konov2} and is difficult to investigate.
The recent progress in the observational low-frequency radio
astronomy, in particular, construction of the LOFAR telescope,
seems very promising as to the thorough studies of pulsar radio
emission at the lowest radio frequencies \citep{stappers}.

The characteristic frequencies of the scattered radiation are the
same as those of the synchrotron re-emission of the particle and
fall into the optical or soft X-ray band. The scattering can
noticeably contribute to the pulsar high-energy emission.
Generally speaking, the mechanism of pulsar high-energy emission
is still a matter of debate \citep[see, e.g.,][for a
review]{harding}. It should be noted that the distinctive feature
of the mechanisms based on synchrotron re-emission of the
particles participating in resonant absorption of the radio
emission \citep{p03,hu05} is a physical connection between the
radio and high-energy emissions of a pulsar. An evidence of such a
connection has recently been found in observations \citep{l07}.
The low-frequency scattering off the gyrating particles is the
additional mechanism of the pulsar high-energy emission, which
also implies a connection with the lowest radio frequencies. Thus,
the scattering process studied in the present paper may have
important observational consequences.

The scattering cross-section derived in Sect.~2.2 is believed to
allow a number of other astrophysical applications. The
spontaneous scattering considered above is expected to be
accompanied with the induced one. Besides that, in the
magnetosphere of a pulsar, there may be other scattering sites,
e.g., in the region of closed magnetic field lines. The scattering
regime examined may also be applicable to synchrotron sources. In
the classical formulation of the problem on the Compton losses in
a synchrotron source, the frequencies of synchrotron emission
greatly exceed the particle gyrofrequency and the assumption of
the scattering by the rectilinearly moving particles is well
justified. However, in case of a substantially broad distribution
function of the particles and/or significant magnetic field
gradients the values of the gyrofrequency lie over a wide range
and the condition of the low-frequency scattering may also be
satisfied. Since this process is much more efficient, it may have
important implications.

\section*{Acknowledgments}

I am grateful to the anonymous referee for useful comments.

\appendix
\section{Derivation of the formulas given by equation (20)}
Below we derive the formulas given by equation (20) in the main
text. To begin with, we apply the induction method to the
well-known Schott formula
\begin{equation}
s(\varepsilon)\equiv\sum_{\nu=1}^\infty\nu^2J_{2\nu}(2\nu\varepsilon)=
\frac{\varepsilon^2(1+\varepsilon^2)}{2(1-\varepsilon^2)^4}.
\label{a1}
\end{equation}
Performing term-by-term differentiation of the series and finding
the first- and second-order derivatives, one can obtain
\begin{equation}
4(1-1/\varepsilon^2)\sum_{\nu=1}^\infty\nu^4J_{2\nu}(2\nu\varepsilon)=
s^{\prime\prime}(\varepsilon)+s^\prime(\varepsilon)/\varepsilon
.\label{a2}
\end{equation}
Above we have taken into account the Bessel equation,
$J_\nu^{\prime\prime}(z)+J_\nu^\prime (z)/z+J_\nu
(1-\nu^2/z^2)=0$. Using equations (\ref{a1})-(\ref{a2}) yields
\begin{equation}
\sum_{\nu=1}^\infty\nu^4J_{2\nu}
(2\nu\varepsilon)=\frac{\varepsilon^2(1+14\varepsilon^2
+21\varepsilon^4+4\varepsilon^6)}{2(1-\varepsilon^2)^7}.\label{a3}
\end{equation}
Then making use of the integral \citep{gr}
\begin{equation}
\int\limits_0^{\pi/2}J_{2\nu}(2\nu x\sin\theta){\rm d}\theta
=\frac{\pi}{2}J_\nu^2(\nu x),\label{a4}
\end{equation}
one can write
\begin{equation}
\sum_{\nu=1}^\infty\nu^4J_\nu^2(\nu x)
=\frac{1}{\pi}\int\limits_0^{\pi/2}x^2\sin^2\theta \frac{
1+14x^2\sin^2\theta+21x^4\sin^4\theta+4x^6\sin^6\theta}
{(1-x^2\sin^2\theta)^7}{\rm d}\theta.\label{a5}
\end{equation}
Performing routine integration, we find finally:
\begin{equation}
\sum_{\nu=1}^\infty\nu^4J_\nu^2(\nu
x)=\frac{x^2(64+592x^2+472x^4+27x^6)}{256(1-x^2)^{13/2}}.\label{a6}
\end{equation}

To get the sum of the analogous series
$\sum_{\nu=1}^\infty\nu^4J_\nu^{\prime^2}(\nu x)$ we proceed from
the well-known formula of the theory of synchrotron emission
\begin{equation}
s(x)\equiv\sum_{\nu=1}^\infty\nu^2J_\nu^2(\nu
x)=\frac{x^2(4+x^2)}{16(1-x^2)^{7/2}}\label{a7}
\end{equation}
and differentiate it twice. This yields
\begin{equation}
2\sum_{\nu=1}^\infty\nu^4J_\nu^{\prime^2}(\nu
x)-2(1-1/x^2)\sum_{\nu=1}^\infty\nu^4J_\nu^4(\nu x)
=s^{\prime\prime}(x)+s^\prime (x)/x,
\end{equation}
which can be reduced to
\begin{equation}
\sum_{\nu=1}^\infty\nu^4J_\nu^{\prime^2}(\nu
x)=\frac{64+624x^2+632x^4+45x^6}{256(1-x^2)^{11/2}}.
\end{equation}


\bsp

\label{lastpage}

\end{document}